\documentclass[showpacs,preprintnumbers,prl,titlepage,preprint,showkeys,nofootinbib,bibnotes,twocolumn,letterpaper,10pt,balancelastpage]{revtex4}
\usepackage{amsmath}
\usepackage{graphicx}
\usepackage{dcolumn}
\usepackage{bm}

\begin{document}

\title{T-Shape Molecular Heat Pump}
\author{Wei-Rong Zhong$^{1,\text{ }2}$}
\email{wrzhong@hotmail.com}
\author{Bambi Hu$^{2,\text{ }3}$}
\affiliation{$^{1}$\textit{Department of Physics, College of Science and Engineering,
Jinan University, Guangzhou, 510632, China.}}
\affiliation{$^{2}$\textit{Department of Physics, Centre for Nonlinear Studies, and The
Beijing-Hong Kong-Singapore Joint Centre for Nonlinear and Complex Systems
(Hong Kong), Hong Kong Baptist University, Kowloon Tong, Hong Kong.}}
\affiliation{$^{3}$\textit{Department of Physics, University of Houston, Houston, Texas
77204-5005, USA.}}
\date{\today }

\begin{abstract}
We report on the first molecular device of heat pump modeled by a T-shape
Frenkel-Kontorova lattice. The system is a three-terminal device with the
important feature that the heat can be pumped from the low-temperature
region to the high-temperature region through the third terminal. The
pumping action is achieved by applying a stochastic external force that
periodically modulates the atomic temperature. The temperature, the
frequency and the system size dependence of heat pump are briefly discussed.
\end{abstract}

\maketitle

A heat pump is a machine or device that moves heat from a low temperature
heat source to a higher temperature heat sink by applying an external work
that modulates the environment of the system \cite{ASHRAE}. In the last
decade, the nanoscale heat pump attracts more and more attentions due to its
valuable applications \cite{Nitzan}. Recent studies have carefully analyzed
the physical mechanism of molecular heat pump through classical \cite{Dhar}
\cite{Nakagawa} and quantum \cite{Segal2} \cite{Segal1} \cite{YDWei}
methods. Typically, in these schemes a carefully-shape external force \cite%
{Segal2} and a stochastic external force \cite{Segal1} periodically modulate
the levels of the nanoobject, leading to the pumping operation. This process
has been confirmed theoretically to be available in a quantum Kubo
oscillator. In classical systems a Brownian noise can lead to the
reverse-direction movement of a molecular motor \cite{Hanggi} \cite{Broek}.
Therefore, it is also expected that stochastic external force can take heat
continuously away from the cold heat bath and pump it into the hot heat bath.

In this letter, we build a classical prototype pumping device driven by
stochastic external force (\emph{\emph{i}.e.,} thermal noise or\emph{\ }heat
bath). In our model, as shown in Fig.1a, the first particle of two segments,
H1, 2C, connects to two end of segment 1O2 via the junction 1 and 2,
respectively. Segment OP and 1O2 are coupled via the mid-particle of Segment
1O2 and the first particle of segment OP. Each segment is a
Frenkel-Kontorova (FK) lattice. The last particle of segment H1 and 2C are
respectively connected to hot heat bath (high temperature, $T_{H}$) and cold
heat bath (low temperature, $T_{C}$), while the last particle of segment OP
is the control terminal. This T-shape FK model, which had ever been used to
study thermal transistor \cite{BLi2} and thermal logic gate \cite{BLi1}, is
similar to the experimental counterpart by putting a polymer chain or a
nanowire on the top of adsorbed sheet \cite{VPouthier}. Furthermore, a quasi
one-dimensional case of this model is the forked nanowire and nanotube,
\emph{\emph{e}.g., }the T-shape nanowire \cite{ZLWang} and Y-type carbon
nanotube \cite{Cummings}, which stand by much potential experimental
execution.
\begin{figure}[htbp]
\begin{center}\includegraphics[width=8cm,height=6cm]{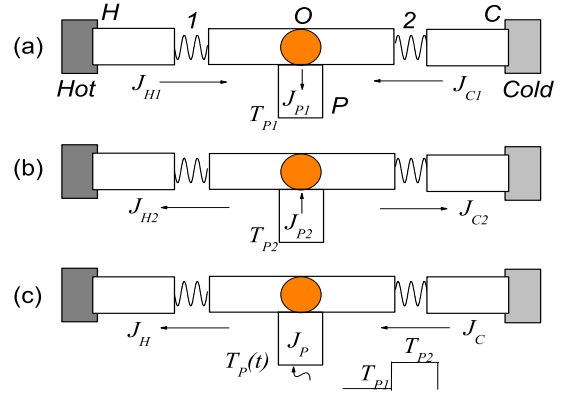}
  \end{center}
  \caption{Configuration of the
heat pump.}
   \label{}
\end{figure}

Segment H1 (2C) and 1O2 are coupled to 1 (2) via a spring of constant $k_{1}$
($k_{2})$. Segments OP and 1O2, which are coupled via the mid-particle of
segment O with a spring of constant $k_{3}$, have the same parameters with
the exception of the size. The number of the particles in segment 1O2 is
nearly twice that in segment OP. The stochastic external force, which is
here presented by a heat bath, is applied to modulate the last particle of
OP segment. The temperature of the heat bath, $T_{P}(t)$, is a periodic
square wave function. As shown in Fig.1c, $T_{P}(t)$ equals to $T_{P1}$ in
the first half period and $T_{P2}$ in the second half period, respectively.
The total Hamiltonian of the model is

\begin{equation}
H=\overset{M}{\sum }H_{M}+\sum H_{int},
\end{equation}%
and the Hamiltonian of each segment can be written as%
\begin{equation}
H_{M}=\sum_{i=1}^{N_{M}}\left[ \frac{p_{M,i}^{2}}{2m_{M}}+\frac{k_{M}}{2}%
\left( x_{M,i+1}-x_{M,i}\right) ^{2}+\frac{V_{M}}{(2\pi )^{2}}\left[ 1-\cos
(2\pi x_{M,i})\right] \right] ,
\end{equation}%
with $x_{M,i}$ and $p_{M,i}$ denote the displacement from equilibrium
position and the conjugate momentum of the $i^{th}$ particle in segment $M$,
where $M$ stands for $H1$, $1O2$, $2C$ or $OP$. $N$ is the number of the
particles in segments $H1$, $2C$ and $OP$. The number of the particles in
segment $1O2$ is $2N+1$. $\sum H_{int}=H_{1}+H_{2}+H_{3}$, in which $%
H_{1}=k_{1}\left( x_{H,1}-x_{O,1}\right) ^{2}/2$, $H_{2}=k_{2}\left(
x_{O,2N+1}-x_{C,1}\right) ^{2}/2$, and $H_{3}=k_{2}\left(
x_{O,N+1}-x_{P,1}\right) ^{2}/2$. We set the masses of all the particles be
unit and use fixed boundaries, $x_{W,N+1}=0,$where W stands for H, C, or P.
The main parameters are $k_{H1}=1.0$, $k_{2C}=9.0$, $k_{1O2}=k_{OP}=3.0$, $%
V_{H1}=2.0$, $V_{2C}=4.5$, $V_{1O2}=V_{OP}=3.0$, $k_{1}=k_{2}=0.05$, $\
k_{3}=3.0$, $T_{H}=0.20$, and $T_{C}=0.16$

In our simulations we use Nose-Hoover thermostat \cite{Nose} and integrate
the equations of motion by using the 4th-order Runge-Kutta algorithm \cite%
{Press}. We have checked that our results do not depend on the particular
thermostat realization (for example, Langevin thermostat). The local
temperature is defined as $T_{i}=\left\langle p_{i}^{2}\right\rangle $, $%
\langle $ $\rangle $ means time average. The local heat flux along the chain
is defined as $J_{W,i}=k_{M}\langle p_{i}(x_{i}-x_{i-1})\rangle $, where $W$
stands for $H$, $C$ and $P$. $i$ is the order of the particle. The heat
current, which flows from $H$ to $C(J_{H})$, or $O$ to $C(J_{C})$, or $O$ to
$P(J_{P})$, is defined as the positive current. The average kinetic energy
is $E_{K}=\sum_{i=1}^{4N+1}\langle v_{i}^{2}\rangle /(4N+1)$, where $v_{i}$
is the velocity of the $i^{th}$ particle and $4N+1$ is the system size,
respectively. The simulations are performed long enough to allow the system
to reach a steady state in which the local heat flux is constant along the
chain.

Here the external perturbation has the same effect as a heat bath coupled to
the particle in the end of segment $OP$. In our model, Figure 2 shows the
temperature of mid-particle of segment $1O2$, $T_{O}$, changes linearly with
the temperature of the heat bath acted on segment P. Due to ballistic
transport of weak link systems in low temperature\textit{\ }\cite{WRZhong1},
the temperature of particle $O$ nearly equals to the temperature of the heat
bath $T_{P}(t)$ \emph{\emph{i}.e., }$T_{O}\approx T_{P}$. As the temperature
of the heat bath connects to P is low enough, the temperature of interface
particle $O$ ($T_{O}$) is also small. Therefore, when $T_{O}$ is smaller
than $T_{H}$ and $T_{C}$, the system absorbs energy from both heat baths and
then the direction of the heat current is shown in Fig.1a, here we set the
heat current $J_{H1}>0$, $J_{C1}<0$, and $J_{P1}>0$. When $T_{O}$ is larger
than $T_{H}$ and $T_{C}$, the energy will dissipate from the system
oscillator mode to heat baths and then the direction of the heat current is
shown in Fig.1b, here the heat current $J_{H2}<0$, $J_{C2}>0$, and $J_{P2}<0$%
. The temperature profiles along the configuration of the system is shown in
the inset of Fig.2 for $T_{P}=0.02$ and $0.22$. Actually, the temperature of
the controlling heat bath is variable. Provided that we use the appropriate
values of $T_{P1}$ and $T_{P2}$ and perform the simulation in long time
enough (the simulation time is far larger than the relaxation time of the
system), as shown in Fig.1c, we will get a negative value of the total heat
current in one period, \emph{\emph{i}.e.,} $J_{H}<0$ and $J_{C}<0$, which
means a pumping operation.
\begin{figure}[htbp]
\begin{center}\includegraphics[width=8cm,height=6cm]{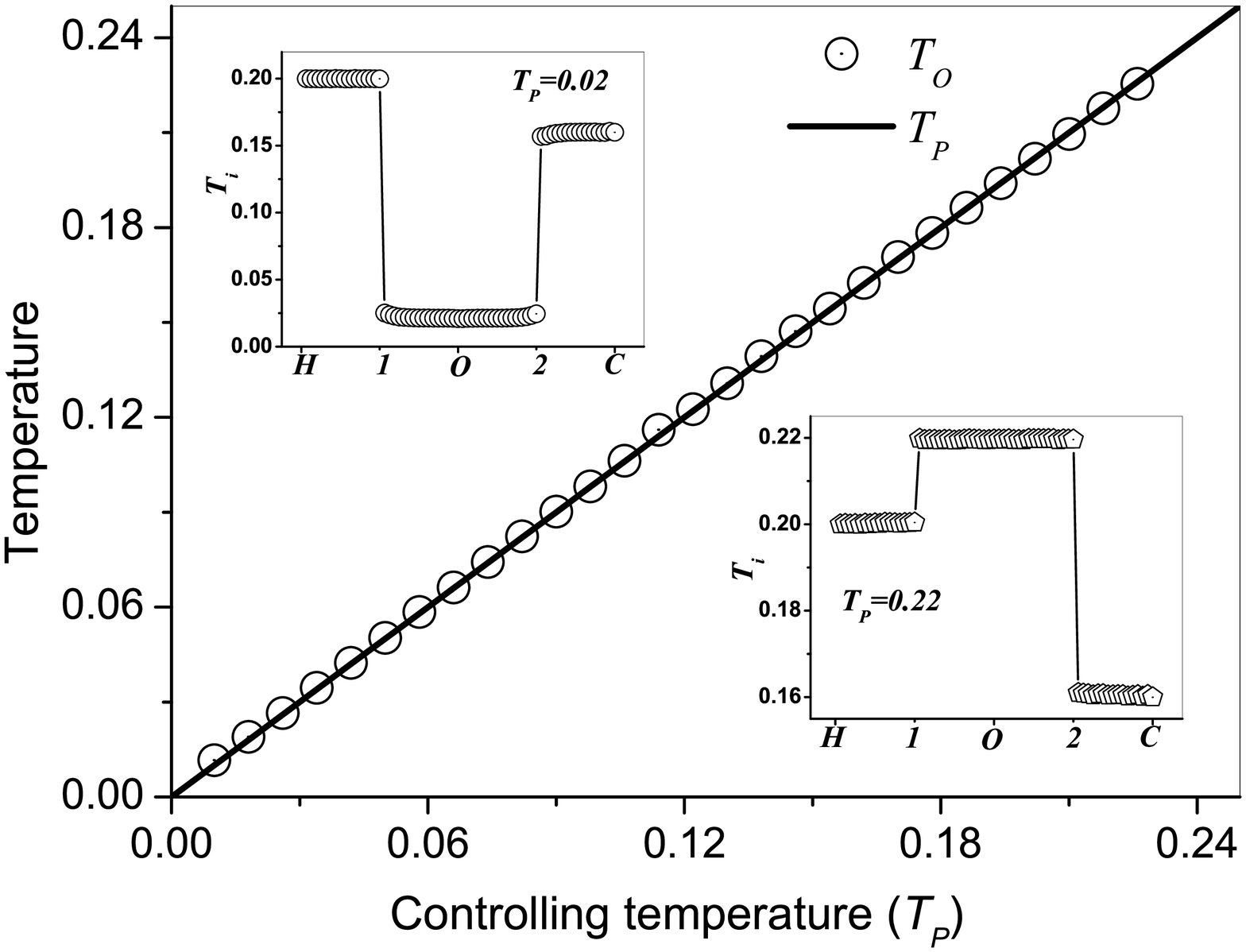}
  \end{center}
  \caption{Relationship between the
temperature of particle $O$ and the temperature of the controlling heat bath
$T_{P}$. The inset is the temperature profiles along segments $H-1-O-2-C$ as
shown in Fig.1a for $T_{P}=0.02$ and $0.22$. }
   \label{}
\end{figure}

This phenomenon can be understood from two essential physical principles:
\textit{negative differential thermal resistance} \textit{(NDTR)} and
\textit{thermal rectification (TR) }\cite{BLi1} \cite{BLi2} \cite{WRZhong1}.
These two effects produce nonlinear relationship between the heat current
and the temperature difference. It can be explained in detail as follows:

As shown in Figs.3a, in the region of low temperature ($T_{O}<T_{C}<T_{H})$,
when temperature $T_{O}$ is decreased by decreasing the temperature of the
controlling heat bath $T_{P}$, $J_{H}$ increases firstly and then decreases
to a small value, however, $J_{C}$ decreases linearly. The dependence of $%
J_{H}$ on the temperature difference $T_{H}-T_{O}$ shows a NDTR effect. In
high temperature region ($T_{O}>T_{H}>T_{C})$, $J_{H}$ and $J_{C}$ have
linear relationships with the temperature. In a brief, segment $HO$ is in
the open state for $T_{O}>T_{H}$ but in the close state for $T_{O}<T_{H}$.
This effect is not visible in segment $OC$. If we select two appropriate
values of $T_{P}$, $T_{P1}=0.02$ and $T_{P2}=0.22$, as the temperature in
the first and the second half period, the total heat currents $%
J_{H}(J_{H1}+J_{H2})$ and $J_{C}(J_{C1}+J_{C2})$ equal to $-4.30\times
10^{-5}$ and $-1.15\times 10^{-5}$, respectively. Heat pump works on the
condition\ that the negative heat flux is larger than the positive one
during a period.
\begin{figure}[htbp]
\begin{center}\includegraphics[width=8cm,height=6cm]{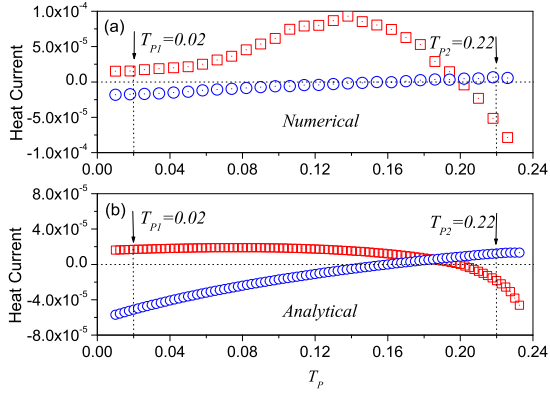}
  \end{center}
  \caption{Heat currents
through two terminals H ($J_{H}$, red square) and C ($J_{C}$, blue circle)
versus the temperature of the controlling heat bath $T_{P}$. (a) Numerical
results for the system size 4N+1=65 and (b) analytical results based on
Eq.(5).  }
   \label{}
\end{figure}

The corresponding analytical method is also included. As reported in Refs
\cite{BHu} and \cite{DHe}, we replace the first and second derivatives of
the external potential by their thermal average with respect to the
effective harmonic Hamiltonian, and then equation 2 can be approximated by
self-consistent phonon theory as

\begin{equation}
\overset{\symbol{126}}{H}_{M}=\sum_{i=1}^{N_{M}}\left[ \frac{p_{M,i}^{2}}{2}+%
\frac{k_{M}}{2}\left( q_{M,i+1}-q_{M,i}\right) ^{2}+\frac{g_{M}}{2}%
(q_{M,i})^{2}\right] ,
\end{equation}%
in which

\begin{equation}
g_{M}=\frac{V_{M}}{2}\exp \left[ -\frac{2\pi T_{M}}{\sqrt{g_{M}(4k_{M}+g_{M})%
}}\right] ,
\end{equation}%
where $T_{M}$ means the average temperature and $M$ refers to segments $HO$,
$OC$ and $OP$. Here we solve the transcendental equation 4 through
calculating the intersection point of left part and right part of the
equation \cite{WRZhong2}. As $k_{1}\longrightarrow 0,$ considering classical
Landauer-type equation, we can get the heat current flows from $H$ to $O$ as

\begin{equation}
J_{H}=\frac{k_{B}(T_{H}-T_{O})}{2\pi }\int_{\omega _{1}}^{\omega _{2}}\chi
(\omega )d\omega ,
\end{equation}%
in which the transmission coefficient is

\begin{equation}
\chi (\omega )\approx \frac{k_{1}^{2}}{k_{H}k_{O}}\sqrt{\frac{%
(4k_{H}+g_{H}-\omega ^{2})(4k_{O}+g_{O}-\omega ^{2})}{(\omega
^{2}-g_{H})(\omega ^{2}-g_{O})}}.
\end{equation}%
\ The cutoff frequencies range from $\omega _{1}=\max \{\sqrt{g_{H}},\sqrt{%
g_{O}}\}$ to $\omega _{2}=\min \{\sqrt{4k_{H}+g_{H}},\sqrt{4k_{O}+g_{O}}\}$,
which correspond to the boundaries of the overlap band of left and right
phonon spectra. Provided that $T_{H}$ and $T_{O}$ are available, then we can
obtain the heat current flows from $H$ to $O$, $J_{H}$. Similarly, as $%
k_{2}\longrightarrow 0$, we can also get the heat current from $O$ to $C$, $%
J_{C}$. Due to the linear relationship of $T_{O}$ and $T_{P}$, figure 3b
analytically confirms the NDTR and TR effects in T-shape FK lattices, which
is numerically presented in Fig.3a.

Heat pump is a dynamical and nonequilibrium effect during macroscopic time.
Therefore, heat pump has some valid conditions. We give three main
parameters which influence the state of heat pump significantly.

\textit{Temperature dependence} In order to obtain four heat currents, $%
J_{H1}$, $J_{H2}$, $J_{C1}$, and $J_{C2}$, shown in Figs.1a and 1b,
obviously the temperature $T_{O}$ should satisfy $T_{O}<T_{C}<T_{H}$ in the
first half period and $T_{C}<T_{H}<T_{O}$ in the second half period,
respectively. Since $T_{O}$ almost equals to $T_{P}$, as illustrated in
Fig.2, $T_{P}$ has to satisfy $T_{P}<T_{C}<T_{H}$ in the first half period
and $T_{C}<T_{H}<T_{P}$ in the second half period, respectively. Here the
main point is what is the range of $T_{P}(t)$. Generally, we fix the low
temperature level $T_{P1}(=0.02)$ and change the high temperature level $%
T_{P2}$. Figure 4 displays that the total heat currents, $J_{H}$ and $J_{C}$%
, are negative as $T_{P2}$ ranges from 0.210 to 0.303. This range of $T_{P2}$
is defined as pumping region. When $T_{P2}<0.210$ or $T_{P2}>0.303$, pumping
effect cannot be realized. It is worth mentioning a point, where $T_{P2}=0.251$ 
and $J_{H}=J_{C}=-0.51\times 10^{-5}<0$, indicates the optimum
pumping.
\begin{figure}[htbp]
\begin{center}\includegraphics[width=8cm,height=6cm]{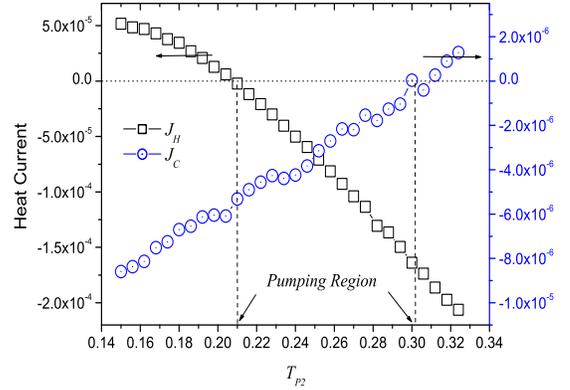}
  \end{center}
  \caption{Temperature dependence of
the heat current in a T-shape FK lattice. The vibration period of $T_{P}(t)$
is $5.0\times 10^{6}$ steps and the system size is 65.}
   \label{}
\end{figure}

\textit{Frequency dependence }In our simulation, the system relaxation time (%
$\tau _{s}$) of our model is about $6.0\times 10^{9}$ simulation steps, then $\tau _{s}$=$6.0\times 10^{7}$. The
oscillating frequency of single particle ($\omega _{p}$, or 1/$\tau _{p}$)
ranges from $2.5\times 10^{-3}$ to $6.2\times 10^{-1}$, which can be
calculated through the phonon spectral analysis of the particle's velocity
\cite{BLi2}. In order to get a stable heat current, the vibration period of
temperature level $T_{P}(t)$ should be far smaller than the system
relaxation time. Furthermore, the vibration period of noise level cannot be
near the oscillating period of single particle. As shown in Fig.5a, in the
case of low frequency, the heat pump works normally with a stable negative
heat current $J_{H}$. However, when the vibration frequency of temperature
is increased to a value larger than $3.0\times 10^{-3}$, the direction of
heat current changes and the heat pump stops working. It is easily
understood from the change of the average kinetic energy with the frequency.
As shown in Fig.5b, the system maintains its average kinetic energy onto a
value $E_{K}$ ($=\frac{1}{2}(E(T_{P1})+E(T_{P2}))=0.072$) at low frequency.
However, the average kinetic energy has a significant change in the
frequency region of single particle, which corresponds to the shadow area in
Fig.5. In this region, the frequency may match the oscillating frequency of
single particle, the temperature of the controlling heat bath has a
significant influence on the oscillating energy of every particle and the
system is nonconservative. Therefore, we set the vibration period of
temperature level, $\tau _{n}$, be $5.0\times 10^{4}$ ,
which satisfies $\tau _{p}<<\tau _{n}<<\tau _{s}$.
\begin{figure}[htbp]
\begin{center}\includegraphics[width=8cm,height=6cm]{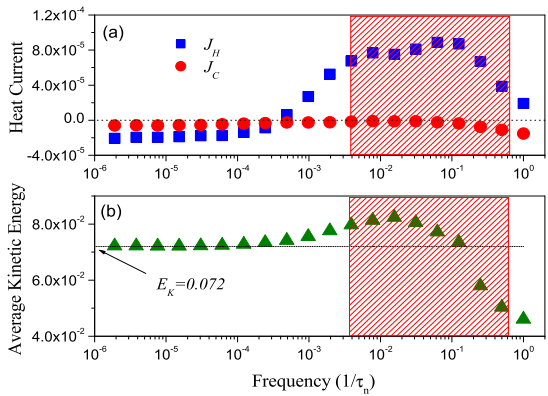}
  \end{center}
  \caption{Frequency dependence of (a) heat currents and (b)
average kinetic energy of heat pump. The remaining parameters are the same
as for Fig.4.}
   \label{}
\end{figure}

\textit{System size effect }The results displayed above are for a system
with 65(4N+1) particles. Since the heat pump mechanism in our model is due
to the coupling between two asymmetric lattices it is reasonable to expect
that the system size will definitely influence the heat pump efficiency. As
shown in Fig.6, the heat current is dependent on the system size. The heat
current of heat pump increases firstly and then decreases by increasing the
number of particles. Finally, the heat pump stops working when the system
size is larger than 500. This phenomenon can be understood by system size
dependence of NDTR \cite{WRZhong1}. When the system size is increased, the
system goes to completely diffusive transport regime and then NDTR
disappears. Thus, the valid condition for heat pump will be unavailable.%

\begin{figure}[htbp]
\begin{center}\includegraphics[width=8cm,height=6cm]{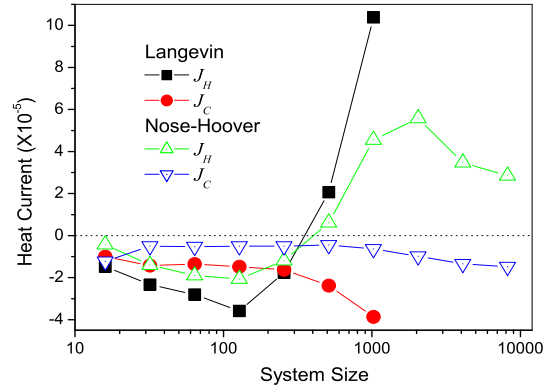}
  \end{center}
  \caption{The finite size effect of the
pumping efficiency for two kinds of heat baths, Langevin (solid Squares and
Circles) and Nose-Hoover (open Up Triangles and Down Triangles). The
remaining parameters are $T_{P1}=0.02$, $T_{P2}=0.22$, and $\protect\tau %
_{n}=5.0\times 10^{6}$.}
   \label{}
\end{figure}

We have to point out that in this letter we use the more complex four
segments rather than three segments FK lattices just for the convenience of
theoretical analysis. Actually, in the case of three segment lattices, the
heat pump may work more efficiently. Finally, we would like to discuss the
improvement of our heat pump model. Figure 1 displays a single heat pump
working between two heat baths with small temperature difference. Moreover,
this kind of single heat pump works with low pumping efficiency. Therefore,
if we expect to get a more powerful heat pump, we can connect single heat
pumps in series or in parallel.

Up to now we only pay attention to the behavior of the device as a heat
pump, which requires $J_{H}<0$ and $J_{C}<0$; however, for the situation
with only $J_{C}<0$ (when the system size is larger than 500 as shown in
Fig.6), which would be useful as a refrigerator mode to extract heat from
the cold source, although in this case this heat will not go to the hot
source but will go to the oscillating heat bath; anyway, the ensemble will
globally act as a refrigerator.

In conclusions, we have reported the feasibility to produce molecular heat
pump based on T-shape FK model. This device can pump the heat from the
low-temperature region to the high-temperature region through controlling
the atomic temperature of the third terminal. Although the heat pump
presented here is only an ideal model, it can be easily imitated in
experiment. The study may also be a valuable illumination in fabricating
nanoscale heat pump. Besides, our heat pump model will help deeply
understand the effect of negative differential thermal resistance.

\begin{acknowledgments}
We would like to thank members of the Centre for Nonlinear Studies for
useful discussions. This work was supported in part by grants from grants
from the Jinan University Young Faculty Research Grant YFRG, the Hong Kong
Research Grants Council RGC and the Hong Kong Baptist University Faculty
Research Grant FRG.
\end{acknowledgments}


\begin{thebibliography}{99}
\bibitem{ASHRAE} The Systems and Equipment volume of the ASHRAE Handbook,
ASHRAE, Inc., Atlanta, GA, (2004).

\bibitem{Nitzan} A . Nitzan, Science, \textbf{317}, 759 (2007).

\bibitem{Dhar} R. Marathe, A. M. Jayannavar, and A. Dhar, Phys. Rev. E
\textbf{75}, 030103(R) (2007).

\bibitem{Nakagawa} N. Nakagawa and T. S. Komatsu, Europhys. Lett. \textbf{75}%
, 22 (2006).

\bibitem{Segal2} D. Segal and A. Nitzan, Phys. Rev. E \textbf{73}, 026109
(2006).

\bibitem{Segal1} D. Segal, Phys. Rev. Lett \textbf{101}, 260601 (2008).

\bibitem{YDWei} Y. Wei, L.Wan, B. Wang, and J. Wang, Phys. Rev. B \textbf{70}%
, 045418 (2004).

\bibitem{Hanggi} P. Hanggi and F. Marchesoni, Artificial Brownian motors:
Controlling transport on the nanoscale, Rev. Mod. Phys. \textbf{81}, 1--55
(2009).

\bibitem{Broek} M. Van den Broek and C. Van den Broeck, Phys. Rev. Lett.
\textbf{100}, 130601 (2008).

\bibitem{BLi2} B. Li, Lei Wang, and Giulio Casati, Appl. Phys. Lett. \textbf{%
88}, 143501 (2006).

\bibitem{BLi1} L. Wang, and B. W. Li, Phys. Rev. Lett. \textbf{99}, 177208
(2007).

\bibitem{VPouthier} V. Pouthier, J. C. Light, and C. Giraredet, J. Chem.
Phys. \textbf{114}, 4955 (2001).

\bibitem{ZLWang} Z. L. Wang, Z. W. Pan, and Z. R. Dai, Microsc. Microanal.
\textbf{8}, 467-474 (2002).

\bibitem{Cummings} A. Cummings, M. Osman, D. Srivastava, and M. Menon, Phys.
Rev. B \textbf{70}, 115405 (2004).

\bibitem{Nose} S. Nose, J. Chem. Phys. \textbf{81}, 511 (1984); W. G.
Hoover, Phys. Rev.A \textbf{31}, 1695 (1985).

\bibitem{Press} W. H. Press, S. A. Teukolsky, W. T. Vetterling, and B. P.
Flannery, Numerical Recipes (Cambridge University Press, Cambridge, 1992).

\bibitem{WRZhong1} W. R. Zhong, P. Yang, B. Q. Ai, Z. G. Shao, and B. Hu,
Phys. Rev. E \textbf{79}, 050103 (2009).

\bibitem{BHu} B. Hu, D. He, L. Yang, and Y. Zhang, Phys. Rev. E \textbf{74},
060101 (2006).

\bibitem{DHe} D. H. He, Thermal Rectification in One-Dimensional Nonlinear
Systems, (PhD Thesis of Hong Kong Baptist University, Hong Kong, 2008).

\bibitem{WRZhong2} W. R. Zhong, Y. Z. Shao, and Z. H. He, Phys. Rev. E
\textbf{74}, 011916 (2006).
\end{thebibliography}
\end{document}